\documentclass[format=sigconf]{acmart}

\usepackage{url}
\usepackage{threeparttable}
\usepackage{multirow}
\usepackage{soul}             

\usepackage{siunitx}
\DeclareSIUnit{\instructionspersecond}{IPS}

\usepackage[acronyms,nohypertypes={acronym},nomain]{glossaries}
\newacronymstyle{custom-acr}{\glsgenacfmt}{

}
\setacronymstyle{custom-acr}

\usepackage{booktabs} 

\usepackage{tikz}
\usetikzlibrary{arrows,calc,positioning,shadows,shapes}

\usepackage{tikz-timing}
\usetikztiminglibrary{nicetabs}
\usetikztiminglibrary{overlays}

\usepackage{enumitem}

\usepackage{xcolor}

\usepackage{subcaption}

\makeatletter
\providecommand*{\input@path}{}
\g@addto@macro\input@path{{src/}}
\makeatother

\newcommand{\amd}{AMD}
\newcommand{\arm}{ARM}
\newcommand{\atom}{Atom}
\newcommand{\buildroot}{Build\-root}
\newcommand{\cadence}{Ca\-dence}
\newcommand{\cortexa}{Cor\-tex-A}
\newcommand{\epiphany}{Epiphany}
\newcommand{\gcc}{GCC}
\newcommand{\intel}{Intel}

\newcommand{\kalraymppa}{Kal\-ray MP\-PA}
\newcommand{\kilocore}{Kilo\-Core}
\newcommand{\kintex}{Kin\-tex-7}
\newcommand{\linaro}{Li\-na\-ro}
\newcommand{\linux}{Linux}
\newcommand{\lowrisc}{lowRISC}
\newcommand{\mentorgraphics}{Men\-tor\-Graph\-ics}
\newcommand{\openembedded}{Open\-Em\-bed\-ded}
\newcommand{\openmp}{OpenMP}
\newcommand{\openpiton}{Open\-Pi\-ton}

\newcommand{\palladium}{Pal\-la\-dium}
\newcommand{\pentium}{Pen\-ti\-um}
\newcommand{\ramp}{RAMP}
\newcommand{\riscv}{\mbox{RISC-V}}
\newcommand{\sthorm}{STHORM}

\newcommand{\ultrascaleplus}{Ultra\-Scale+}
\newcommand{\veloce}{Ve\-lo\-ce}
\newcommand{\virtex}{Vir\-tex-7}
\newcommand{\vivado}{Vi\-va\-do}
\newcommand{\xeon}{Xeon}
\newcommand{\xilinx}{Xilinx}
\newcommand{\zynq}{Zynq}

\newcommand{\todo}[1]{}

\newcommand{\glsknown}[1]{\glsunset{#1}\gls{#1}}
\newcommand{\glsknownpl}[1]{\glsunset{#1}\glspl{#1}}
\newcommand{\Glsknown}[1]{\glsunset{#1}\Gls{#1}}

\newcommand{\glsunknown}[1]{\glsreset{#1}\gls{#1}}
\newcommand{\glsunknownpl}[1]{\glsreset{#1}\glspl{#1}}
\newcommand{\Glsunknown}[1]{\glsreset{#1}\Gls{#1}}
\newcommand{\Glsunknownpl}[1]{\glsreset{#1}\Glspl{#1}}

\newcommand{\secref}[1]{\S\,\ref{sec:#1}}
\newcommand{\Figref}[1]{Fig.~\ref{fig:#1}}
\newcommand{\figref}[1]{\Figref{#1}}

\acmConference[CARRV'17]{Computer Architecture Research with RISC-V Workshop}{October 14, 2017}{Boston, MA, USA}
\setcopyright{none}

\begin{document}

\definecolor{sig0color}{HTML}{90c7ec} 
\definecolor{sig1color}{HTML}{ffb97d} 

\newcommand{\workhandle}{HERO}
\title{\workhandle: Heterogeneous Embedded Research Platform~\\for Exploring RISC-V Manycore Accelerators on FPGA}
\renewcommand{\shorttitle}{\workhandle: Heterogeneous Embedded Research Platform for Exploring RISC-V PMCAs}

\author{Andreas Kurth}
\orcid{0000-0001-5613-9544}
\email{kurth@iis.ee.ethz.ch}
\author{Pirmin Vogel}
\email{vogel@iis.ee.ethz.ch}
\affiliation{%
  \institution{Integrated Systems Laboratory, ETH Zurich}
}
\author{Alessandro Capotondi}
\email{alessandro.capotondi@unibo.it}
\affiliation{%
  \institution{Microelectronics Research Group, University of Bologna}
}
\author{Andrea Marongiu}
\email{marongiu@iis.ee.ethz.ch}
\author{Luca Benini}
\email{benini@iis.ee.ethz.ch}
\affiliation{%
  \institution{Integrated Systems Laboratory, ETH Zurich}
}
\affiliation{%
  \institution{Microelectronics Research Group, University of Bologna}
}

\renewcommand{\shortauthors}{A.\ Kurth, P.\ Vogel, A.\ Capotondi, A.\ Marongiu, and L.\ Benini}

\newacronym{acp}{ACP}{Accelerator Coherency Port}
\newacronym{api}{API}{application programming interface}
\newacronym{apu}{APU}{auxiliary processing unit}
\newacronym{asic}{ASIC}{application specific integrated circuit}
\newacronym{axi}{AXI}{Advanced eXtensible Interface}

\newacronym{bram}{BRAM}{block RAM}

\newacronym{cpu}{CPU}{central processing unit}

\newacronym{dma}{DMA}{direct memory access}
\newacronym{dram}{DRAM}{dynamic random access memory}
\newacronym{dsp}{DSP}{digital signal processing}


\newacronym{fame}{FAME}{FPGA Architecture Model Execution}
\newacronym{farm}{FARM}{Flexible Architecture Research Machine}
\newacronym{fpga}{FPGA}{field-programmable gate array}
\newacronym{fpu}{FPU}{floating-point unit}
\newacronym{ff}{FF}{flip-flop}

\newacronym{gpu}{GPU}{graphics processing unit}

\newacronym[longplural={heterogeneous embedded systems on chip}]{hesoc}{HESoC}{heterogeneous embedded system on chip}
\newacronym{hpc}{HPC}{high performance computing}
\newacronym{hw}{HW}{hardware}

\newacronym{ic}{IC}{integrated circuit}
\newacronym{io}{I/O}{input/output}
\newacronym{iot}{IoT}{internet of things}
\newacronym{ip}{IP}{intellectual property}
\newacronym{isa}{ISA}{instruction set architecture}

\newacronym{juno}{Juno ADP}{Juno ARM Development Platform}


\newacronym[shortplural={LDSes}]{lds}{LDS}{linked data structure}
\newacronym{lsb}{LSB}{least significant bit}
\newacronym{lto}{LTO}{link-time optimization}
\newacronym{lut}{LUT}{lookup table}

\newacronym{mmu}{MMU}{memory management unit}
\newacronym{mpsoc}{MPSoC}{multiprocessor system on chip}

\newacronym{noc}{NoC}{network on chip}

\newacronym{os}{OS}{operating system}

\newacronym{pe}{PE}{processing element}
\newacronym{pl}{PL}{programmable logic}
\newacronym{pmca}{PMCA}{programmable manycore accelerator}
\newacronym{ptw}{PTW}{page table walk}
\newacronym{pulp}{PULP}{Parallel Ultra Low Power}


\newacronym{rab}{RAB}{remapping address block}
\newacronym{rte}{RTE}{runtime environment}

\newacronym[longplural={systems on chip}]{soc}{SoC}{system on chip}
\newacronym[longplural={scratchpad memories}]{spm}{SPM}{scratchpad memory}
\newacronym{svm}{SVM}{shared virtual memory}
\newacronym{sw}{SW}{software}

\newacronym{tlb}{TLB}{translation lookaside buffer}


\newacronym{va}{VA}{virtual address}
\newacronym{vmm}{VMM}{virtual memory management}




\newacronym{zc706}{ZC706}{Xilinx Zynq ZC706 Evaluation Kit}

\fontsize{8.7}{10.6}\selectfont

\begin{abstract}
  \Glsunknownpl{hesoc} co-integrate a standard host processor with \glsunknownpl{pmca} to combine general-purpose computing with domain-specific, efficient processing capabilities.
  While leading companies successfully advance their \gls{hesoc} products, research lags behind due to the challenges of building a prototyping platform that unites an industry-standard host processor with an open research \gls{pmca} architecture.

  In this work we introduce \workhandle, an \acrshort{fpga}-based research platform that combines a \gls{pmca} composed of clusters of \riscv{} cores, implemented as soft cores on an \acrshort{fpga} fabric, with a hard \arm{} \cortexa{} multicore host processor.
  The \gls{pmca} architecture mapped on the \acrshort{fpga} is silicon-proven, scalable, configurable, and fully modifiable.
  \workhandle{} includes a complete software stack that consists of a heterogeneous cross-compilation toolchain with support for \openmp{} accelerator programming, a \linux{} driver, and runtime libraries for both host and \gls{pmca}.
  \workhandle{} is designed to facilitate rapid exploration on all software and hardware layers:
  run-time behavior can be accurately analyzed by tracing events,
  and modifications can be validated through fully automated hardware and software builds and executed tests.
  We demonstrate the usefulness of \workhandle{} by means of case studies from our research. 
  \glsresetall
\end{abstract}



\thanks{This work was partially funded by the EU's H2020 projects HERCULES~(No.~688860) and OPRECOMP~(No.~732631).}

\maketitle

\setlength{\floatsep}{2pt plus 16pt minus 1pt}  
\setlength{\textfloatsep}{\floatsep}            
\setlength{\intextsep}{\floatsep}               
\setlength{\dblfloatsep}{\floatsep}             
\setlength{\dbltextfloatsep}{\floatsep}         
\setlength{\abovecaptionskip}{8pt plus 10pt}    
\setlength{\belowcaptionskip}{4pt plus 10pt}    

\section{Introduction}

\Glspl{hesoc} are used in various application domains to combine general-purpose computing with domain-specific, efficient processing capabilities.
Such architectures co-integrate a general-purpose host processor with \glspl{pmca}.
While leading companies continue to advance their products~\cite{Anandtech:2017:AppleA10X, AnandTech:2017:SamsungExynos8895, AnandTech:2017:NvidiaTegraParker}, computer architecture research on such systems lags behind: little is known on the internals of these products, and there is no research platform available that unites an industry-standard host processor with a modifiable and extensible \gls{pmca} architecture.

An important aspect of processors is their \gls{isa}, because it is the interface between software and hardware and ultimately determines their usability and performance in the system.
The \riscv{} \gls{isa}~\cite{Waterman:2016:RiscvIsa} has recently gained considerable momentum in the community~\cite{Zimmer:2016:RiscvVectorProcessor, Celio:2015:Boom, Gautschi:2017:NtvRiscvDspExtensions} because it is an open standard and designed in a modular way: a small set of base instructions is accompanied by standard extensions and can be further extended through custom instructions~\cite{MicroprocessorReport:2016:Riscv}.
This allows computer architects to implement the extensions suitable for their target application.
Moreover, the \gls{isa} is suitable for various types of processors from tiny microcontrollers~\cite{Traber:2016:Pulpino} to high-performance superscalar out-of-order cores \todo{find reference}, because it does not specify implementation properties.
Combined, these characteristics make \riscv{} an interesting candidate for specialized \glspl{pmca}.

There are many different \gls{pmca} architectures, such as \kalraymppa{}~\cite{DeDinechin:2013:KalrayMPPA}, \kilocore{}~\cite{Bohnenstiehl:2017:KiloCore}, \sthorm{}~\cite{Melpignano:2012:P2012}, \epiphany~\cite{Olofsson:2016:EpiphanyV}, and \glsknown{pulp}~\cite{Rossi:2014:PULP}.
\Gls{pulp} is an architectural template for scalable, energy-efficient processing that combines an explicitly-managed memory hierarchy, \gls{isa} extensions and compiler support for specialized \glsknown{dsp} instructions, and energy-efficient cores operating in parallel to meet processing performance requirements.
\Gls{pulp} is a silicon-proven~\cite{Gautschi:2017:NtvRiscvDspExtensions}, open~\cite{Traber:2016:Pulpino} architecture implementing the \riscv{} \gls{isa} and can cover a wide range of performance requirements by scaling the number of cores or adding domain-specific extensions.
Thus, it is ideally suited to serve as a baseline \gls{pmca} in research on \glspl{hesoc}.

Research on heterogeneous systems traditionally follows a two-pron\-ged approach: hardware accelerators are developed and evaluated in isolation~\cite{Johnson:2011:Rigel, Chen:2015:NnAccelerator}, and their impact on system-level performance is estimated through models and simulators~\cite{Li:2009:McPAT, Bortolotti:2016:VirtualSoC}.
Compared to implementing accelerators in prototype heterogeneous systems, this approach has significant drawbacks, however:
First, interactions between host, accelerators, the memory hierarchy, and peripherals are complex to model accurately, making simulations orders of magnitude slower than running prototypes. 
Second, even full system simulators such as gem5~\cite{Binkert:2011:gem5} model \glspl{hesoc} to a limited degree only~\cite{Butko:2016:bigLITTLESimulation}.
For example, models of system-level interconnects or \glspl{mmu}, which dynamically influence the path from accelerators to different levels of the memory hierarchy, are missing.
Third, simulations are based on assumptions.
Contrary to results obtained through implementation, simulated results burden authors and reviewers with having to justify and validate the underlying assumptions.
Working prototypes, on the other hand, enable efficient, collaborative, and accurate computer architecture research and development, which can compete with industry's pace~\cite{Lee:2016:AgileHardware}.
To perform \emph{system}-level research using standard benchmarks and real-world applications, however, the system must additionally be efficiently programmable: a heterogeneous programming model and support for \gls{svm} between host and \glspl{pmca} are indispensable.


In this work, we present \workhandle{}, the first (to the best of our knowledge) heterogenous manycore research platform.
\workhandle{} combines an \arm{} \cortexa{} host processor with a scalable, configurable, and extensible \glsknown{fpga} implementation of a silicon-proven, cluster-based \gls{pmca}~(\secref{platform_hardware}).
\workhandle{} will be released open-source and includes the following core contributions:
\begin{itemize}[leftmargin=*]
  \item A heterogeneous software stack~(\secref{platform_software}) that supports \openmp{}~4.5 and \gls{svm} for transparent accelerator programming, which tremendously simplifies porting of standard benchmarks and real-world applications and enables system-level research.
  \item Profiling and automated verification solutions that enable efficient hardware and software R\&D on all layers~(\secref{hardware_software_RnD_tools}).
\end{itemize}
With up to \num{64}~\riscv{} cores running at more than \SI{30}{\mega\hertz} on a single \gls{fpga}~(\secref{implemented_platforms}), \workhandle{}'s \gls{pmca} implementation is nominally capable of executing more than \SI{1.9}{\giga\instructionspersecond} and outperforms cycle-accurate simulation by orders of magnitude.
We demonstrate \workhandle{}'s capabilities by means of case studies from our research~(\secref{parallelization_evaluation} to \secref{event_tracing_and_analysis_evaluation}).
\section{Platform}
\label{sec:platform}

In this section, we present first the hardware and then the software infrastructure of our research platform.

\begin{figure*}[ht]
  \centering
  \includegraphics[width=\textwidth]{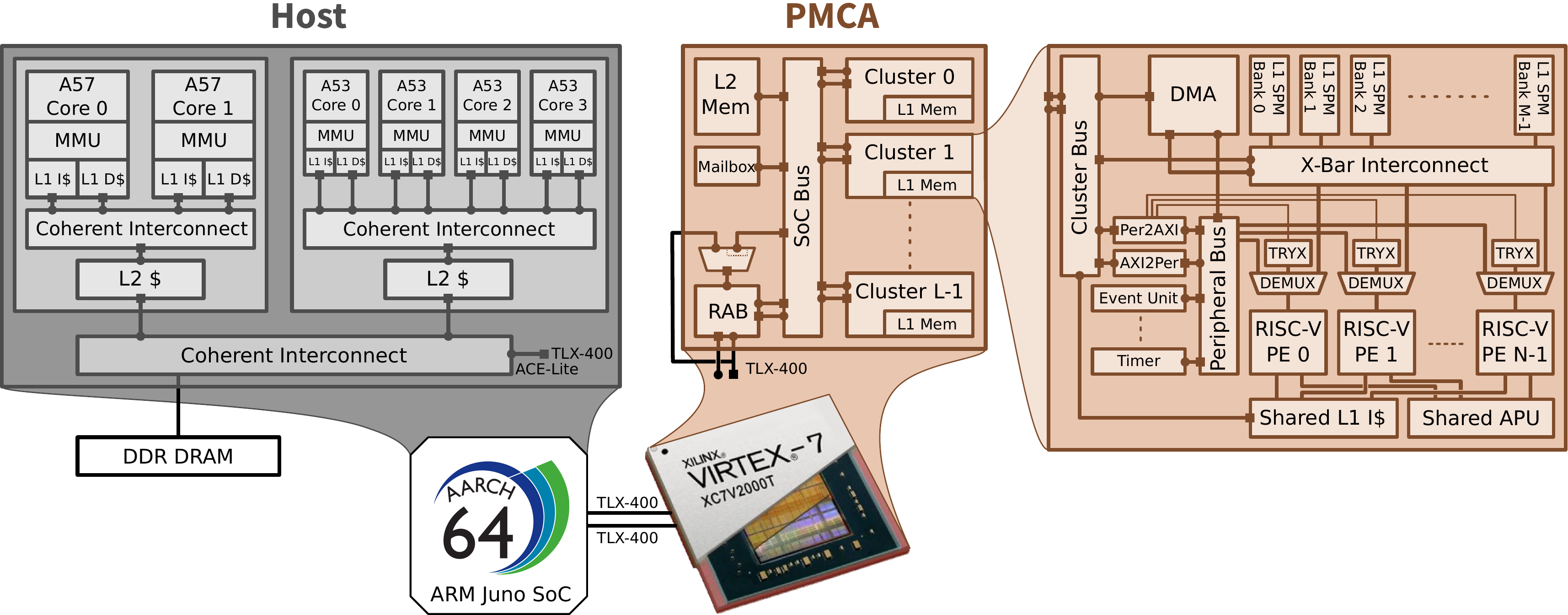}
  \caption{\workhandle{}'s hardware, as implemented on the \acrshort{juno}.}
  \label{fig:juno_pmca}
\end{figure*}
\begin{figure}[ht]
  \centering
  \includegraphics[width=.7\columnwidth]{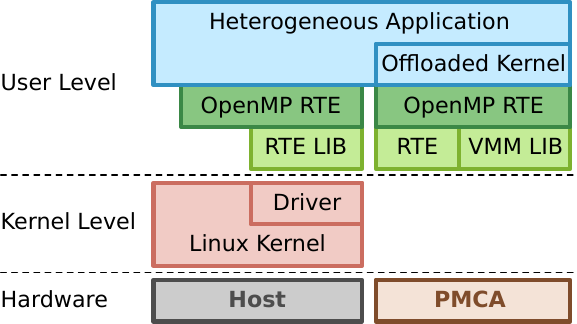}\\[.5\baselineskip]
  \caption{\workhandle{}'s software stack.}
  \label{fig:software_stack}
\end{figure}

\subsection{Hardware}
\label{sec:platform_hardware}

\workhandle{} can be implemented on different hardware platforms consisting of a hard-\glsknown{ip}, \arm{} \cortexa{} host \glsknown{cpu} and an \gls{fpga} fabric used to implement the \gls{pmca}.
Fig.~\ref{fig:juno_pmca} gives an overview of the implementation of \workhandle{} on the \gls{juno}, which will be discussed in detail in \secref{implemented_platforms}.
On all platforms, logic instantiated in the \gls{fpga} can access the shared main \gls{dram} through a low-latency \glsknown{axi} interface coherently with the caches of the host.
This qualifies the platforms for development and prototyping of tightly-integrated accelerators and the associated software infrastructure.

\paragraph{\Glsunknown{pmca}}
As \gls{pmca}, \workhandle{} uses the latest version of the \gls{pulp} platform~\cite{Rossi:2014:PULP}.
\gls{pulp} has been employed in multiple research \glspl{asic} designed for parallel ultra-low power processing.
To overcome scalability limitations, it uses a multi-cluster design and relies on multi-banked, software-managed \glspl{spm} and lightweight, multi-channel \gls{dma} engines instead of data caches.
The 32b \riscv{} \glspl{pe}~\cite{Gautschi:2017:NtvRiscvDspExtensions} within a cluster primarily operate on data present in the shared L1 \gls{spm} to which they connect through a low-latency, logarithmic interconnect.
The \glspl{pe} use the cluster-internal \gls{dma} engine to copy data between the local L1 \gls{spm} and remote \glspl{spm} or shared main memory.
Transactions to main memory pass through the \gls{rab}~\cite{Vogel:2015:RAB}, which performs virtual-to-physical address translation based on the entries of an internal table, similar to the \glspl{mmu} of the host \gls{cpu} cores.
This lightweight hardware block is managed in software directly on the \gls{pmca}~\cite{Vogel:2017:EfficientVirtualMemory}.
The host and the \gls{pmca} can thus efficiently share virtual address pointers.
As such, \gls{svm} substantially eases overall system programmability and enables efficient sharing of linked data structures in the first place.

Since \workhandle{} uses \gls{fpga} logic to implement the \gls{pmca}, it cannot reach the high clock frequency and energy efficiency of an \gls{asic} implementation.
Nevertheless, the performance of a fully integrated \gls{hesoc} can be accurately determined:
One option is to proportionally scale down the clock frequency of host and \gls{dram}.
Even more accurately, the provided tracing infrastructure~(\secref{event_tracing_and_analysis}) can be used to monitor the interfaces of the \gls{pmca}, from which the effect of clock frequency ratio differences between \gls{pmca}, host, and \gls{dram} can be calculated.
The platform is perfectly suited for studying the system-level integration of a \gls{pmca}, developing heterogeneous software, and exploring architectural variations of the \gls{pmca} including, e.g., cluster-internal \glspl{apu} and application-specific accelerators, hardware-managed caches and coherency protocols, interconnect topologies, and scalable system \glspl{mmu}.
In essence, the \gls{pmca} is composed of exchangeable and modifiable blocks and interfaces, and different architectures can be derived from our implementation to match individual research interests.

The \gls{pmca} is highly configurable.
Tab.~\ref{tab:options} gives an overview of the different configurability options currently supported.
Besides the number of clusters and the number of \glspl{pe} and \gls{spm} banks per cluster, the 32b \riscv{} \glspl{pe} themselves can be configured to trade off hardware resources and computing performance.
The single-precision \gls{fpu} can be private, moved to the \gls{apu} to be shared among multiple \glspl{pe} within a cluster or completely disabled.
Similarly, the integer \gls{dsp} extension unit, the divider, and the multiplier can be private or shared in the \gls{apu}.
In addition, different designs for the shared instruction caches (single- or multi-ported) and top-level interconnects (bus or network on chip) can be selected.
Also the design of the \gls{rab} is configurable:
The number of \gls{tlb} entries and levels as well as the architecture of the second-level \gls{tlb} can be adjusted.

\begin{table}[h]
  \centering
  \caption{Configuration options for \workhandle{}'s \acrshort{pmca}}
  \label{tab:options}
  \begin{threeparttable}
    \begingroup
    \setlength{\tabcolsep}{5pt} 
    \begin{tabular}{lr}
    \toprule
    Component & Options \\ \midrule
    \#\,Clusters   & \ul{1}, 2, 4, \textbf{8} \\
    System-level interconnect & \textbf{\ul{Bus}} or network on chip \\ \midrule
    \#\,\glspl{pe} per cluster & 2, 4, \textbf{\ul{8}} \\
    \gls{fpu} & Private, shared (\gls{apu}), \textbf{\ul{off}} \\
    Integer \gls{dsp} unit, divider, multiplier & \textbf{\ul{Private}}, shared (\gls{apu}) \\ \midrule
    L1 \glspl{spm} \#\,banks & 4, 8, \textbf{\ul{16}} \\
    L1 \glspl{spm} size [\si{\kibi\byte}] & 32, 64, 128, \textbf{\ul{256}} \\
    L2 \gls{spm} size [\si{\kibi\byte}] & 32, 64, 128, \textbf{\ul{256}} \\
    Instruction cache design & \textbf{Single-} or \ul{multi-ported} \\
    Instruction cache size [\si{\kibi\byte}] & 2, \ul{4}, \textbf{8} \\
    Instruction cache \#\,banks & 2, \ul{4}, \textbf{8} \\ \midrule
    \gls{rab} L1 \gls{tlb} size & 4, 8, 16, \textbf{\ul{32}}, 64 \\
    \gls{rab} L2 \gls{tlb} size & 0, 256, 512, \textbf{\ul{1024}}, 2048 \\
    \gls{rab} L2 \gls{tlb} associativity & 16, \textbf{\ul{32}}, 64 \\
    \gls{rab} L2 \gls{tlb} \#\,banks & 1, 2, \textbf{\ul{4}}, 8 \\
    \bottomrule
    \end{tabular}
  \endgroup
  \begin{tablenotes}
    \footnotesize
    \item Bold and underlined values refer to implementations discussed in \secref{implemented_platforms}.
  \end{tablenotes}
  \end{threeparttable}
\end{table}

\subsection{Software}
\label{sec:platform_software}

In this section, we describe the various components of \workhandle{}'s software stack.
Fig.~\ref{fig:software_stack} shows how the different software layers and components of host and \gls{pmca} interact.
These components seamlessly integrate the \gls{pmca} into the host system and allow for transparent accelerator programming using \openmp{}~4.5.
The application developer can write and compile a single application source.
Application kernels suitable for offloading to the \gls{pmca} can simply be encapsulated using the \openmp{} \texttt{target} directive.
The actual offload is then taken care of by the \openmp{} \gls{rte}.

\subsubsection{Heterogeneous \texorpdfstring{\openmp{}}{OpenMP} programming}
\label{sec:heterogeneous_OpenMP_programming}
To allow the host \openmp{} \gls{rte} to actually perform an offload to the \gls{pmca}, a custom plugin that contains the \gls{pmca}-specific implementations of the generic \glspl{api} is required~\cite{Capotondi:2017:OpenMPOffloading}.
For example, this plugin defines how the input and output variables specified in the \texttt{target} construct are passed between host and \gls{pmca}.
\workhandle{} currently supports both copy-based and zero-copy offload semantics.
The latter exploits the \gls{svm} capabilities of the platform to just pass virtual address pointers, thereby avoiding costly data copies to a physically contiguous, uncached section in main memory and enabling efficient sharing of linked data structures.

\subsubsection{Heterogeneous cross compilation}
Generating both host and \gls{pmca} binaries from a single application source requires a set of compiler extensions to build a heterogeneous cross-compilation toolchain based on \gcc{}~5.2~\cite{Capotondi:2017:OpenMPOffloading}.
When the compiler expands a \texttt{target} construct in the front end, a new function is outlined that is ultimately also compiled for the \gls{pmca}.
This is achieved by first streaming the functions to be offloaded into a \gls{lto} object, which is then fed to a custom offload tool at link time.
This tool first recompiles the functions for the target \gls{pmca} using the \riscv{} back-end compiler and links all \gls{pmca} runtimes and libraries.
It then creates the hooks for the offloadable functions required by the host \openmp{} runtime.
Finally, the tool packs everything inside a dedicated section in the host binary.

An additional compiler pass is used to instrument all load and stores of the \gls{pmca} to variables in \gls{svm} within the \texttt{target} constructs with calls to low-overhead macros~\cite{Vogel:2016:PointerRich}.
These macros protect the \gls{pmca} from using invalid responses returned by the hardware in case of \gls{tlb} misses in the \gls{rab} and interface with the \gls{vmm} library~\cite{Vogel:2017:EfficientVirtualMemory}.

\subsubsection{Host runtime library and \texorpdfstring{\linux{}}{Linux} driver}
The host \gls{rte} library interfaces the host-side \openmp{} runtime with the \linux{} driver.
In addition, it is used to reserve all virtual addresses overlapping with the physical address map of the \gls{pmca}.
This is required as any access of the \gls{pmca} to a shared variable located at such an address would not be routed to \gls{svm} but instead to its internal \glspl{spm} or memory-mapped registers.
The driver handles low-level tasks such as interrupt handling, synchronization between \gls{pmca} and host, host cache maintenance, operation of the system-level \gls{dma} engine (e.g. to offload the \gls{pmca} binary), operating the profiling hardware, and initially setting up the \gls{rab} to give the \gls{pmca} access to the page table of the heterogeneous user-space application.

\subsubsection{\texorpdfstring{\Acrshort{pmca} \glsunknown{vmm}}{PMCA virtual memory management (VMM)} library}
Having access to the page table of the heterogeneous user-space application, the \gls{pmca} can operate its virtual memory hardware autonomously.
A \gls{vmm} library~\cite{Vogel:2017:EfficientVirtualMemory} on the \gls{pmca} abstracts away differences between host architectures and \gls{rab} configurations and provides a uniform \gls{api} to explicitly map pages and handle \gls{rab} misses.
When a core accesses virtual memory through the \gls{rab}, the corresponding address translation may not be configured.
In this case, the core that caused the miss goes to sleep and the miss is enqueued in the \gls{rab}.
To handle a miss, the \gls{vmm} library dequeues it, translates its virtual address to a physical one by walking the page table of the host user-space process, selects a \gls{rab} table entry to replace and configures it accordingly, and wakes up the core that caused the miss.
The \gls{vmm} library is compatible with any host architecture supported by the \linux{} kernel.

\subsection{Tools for Hardware and Software R\&D}
\label{sec:hardware_software_RnD_tools}

\subsubsection{Event tracing and analysis}
\label{sec:event_tracing_and_analysis}

Fine-grained information on the run-time behavior of the \gls{pmca} in the \gls{hesoc} is crucial for both hardware and software engineers to evaluate their designs and implementations.
While simulations can provide first estimates, they do not accurately reproduce run-time behavior of \glspl{hesoc}, as stated in the introduction.
Instead, this information can be extracted by \emph{tracing events} in the running \gls{hesoc} prototype, which poses the following challenges:
First, the tracer must not interfere with program execution; in particular, inserting instructions (e.g., to write memory) is not an option.
Second, the tracer must be cycle-accurate to allow analysis of rapid, consecutive cause-effect events yet be able to handle measurements spanning millions of events to cover complex applications.
Third, the tracer should use \gls{fpga} resources economically to not hamper the evaluation of complex hardware.
Fourth, the tracer should not require modifications of applications, but should allow application-specific analyses.

\workhandle{}'s event tracing solution is a hybrid design composed of lightweight tracer hardware blocks, which can be inserted anywhere on the \gls{fpga}, and a driver on the host.
The customizable tracer hardware blocks are attached to signals and record their values as timestamped event when user-specified activation conditions are met.
They store events in dedicated, local buffers implemented with \glspl{bram}.
When a buffer is full, the tracer hardware stops the \gls{pmca} by disabling its clock and raises an interrupt to delegate control to a driver on the host.
The driver then reads out all events from the buffers to main memory, clears the buffers, and re-enables the clock of the \gls{pmca}.
This process is entirely transparent to the \gls{pmca}, whose state is frozen during the transfer. 
The timestamps of all loggers are synchronized because they are driven from a common clock, which is disabled with the \gls{pmca} clock.

After an application terminated, all traced events are available in main memory for analysis.
\workhandle{}'s event analysis software processes the data in three layers.
The first layer is generic: it reads the binary data from memory and creates a time-sorted list of events, which contain generic meta data and the ID of the tracer, and a collection of properties of the platform on which they were recorded.
The second layer is measurement- and platform-specific.
For example, if a logger traced memory accesses by cores, each of its generic events is converted to a read/write access to a memory address by one core; if a logger traced synchronization events, each is converted to a set of involved cores.
The third layer is application-specific and optional: by linking run-time information such as memory accesses and synchronization events with knowledge about algorithms and data structures, questions about bottlenecks and how hardware and software design choices affect them can be answered very precisely.


\subsubsection{Automated Implementation and Validation}

\begin{figure}[ht]
  \centering
  \tikzstyle{hw} = [fill=gray!45]
  \tikzstyle{sw} = [fill=gray!15]
  \tikzstyle{block} = [rectangle, draw, fill=white, inner sep=2pt, align=center]
  \tikzstyle{action} = [block, text width=6em, rounded corners=1mm]
  \tikzstyle{automatic} = [draw, -latex']
  \tikzstyle{manual} = [automatic, densely dotted]
  \begin{tikzpicture}[font=\scriptsize, auto, node distance=3.5mm, outer sep=1pt]
    \node [action, hw] (hdl) {change HDL};
    \node [coordinate, left=of hdl] (hdl entry) {};
    \node [action, hw, right=of hdl] (unit test) {simulate unit test};
    \node [action, hw, right=of unit test] (hdl commit) {commit HDL change};
    \node [action, hw, below left=3.25mm and -9mm of hdl commit, copy shadow] (sim tbs) {simulate all TBs};
    \node [action, hw, below right=2mm and -9mm of hdl commit, copy shadow] (bitstreams) {build bitstreams for all targets};
    \node [action, hw, below=11mm of hdl commit, copy shadow] (check and deploy) {check results and deploy bitstreams};
    \node [block, left=of check and deploy, text width=4em, minimum height=4em, copy shadow] (target platforms) {target platforms};
    \node [action, sw, below=of check and deploy, text width=9em, double copy shadow] (execute) {execute all\\(app., build conf., run param.)\\combinations on all platforms};
    \node [action, sw, below=of execute, copy shadow] (build sw) {build all tests and benchmarks in all (platform-specific) configs.};
    \node [action, sw, left=of build sw, yshift=-3mm] (sw commit) {commit SW change};
    \node [action, sw, above=of sw commit] (individual tests) {compile and run individual tests};
    \node [action, sw, left=of individual tests] (sw) {change SW};
    \node [coordinate, left=of sw] (sw entry) {};
    \draw [manual] (hdl entry) -- (hdl);
    \draw [manual] (hdl) -- (unit test);
    \draw [manual] (unit test) -- (hdl commit);
    \draw [manual] (unit test.north) -- ++(0,3mm) -| node[pos=0.25, above, yshift=-3pt] {\textit{if failed}} (hdl);
    \draw [automatic] (hdl commit) -- (sim tbs);
    \draw [automatic] (hdl commit) -- (bitstreams);
    \draw [automatic] (bitstreams) -- (check and deploy);
    \draw [automatic] (sim tbs) -- (check and deploy);
    \draw [automatic] (check and deploy) -- (target platforms);
    \draw [automatic] (check and deploy) -- (execute);
    \draw [automatic] (execute) -| ($(target platforms.south) + (1mm,0) $);
    \draw [automatic] (build sw) -- (execute);
    \draw [automatic] (sw commit) -- (build sw.west |- sw commit);
    \draw [manual] (individual tests) -- (sw commit);
    \draw [manual] (sw) -- (individual tests);
    \node [coordinate] (individual tests exec oup) at ($ (individual tests.north) + (1mm,0) $) {};
    \draw [manual] (individual tests exec oup) -- node[midway, above, sloped, yshift=-2pt] {\textit{execute}} (target platforms.south -| individual tests exec oup);
    \draw [manual] (sw entry) -- (sw);
    \draw [manual] (individual tests.north) ++(-2mm,0) -- ++(0,2mm) -| node[pos=0.25, above, yshift=-3pt] {\textit{if failed}} (sw);
    \node [below right=5mm and 5mm of hdl entry] (auto label) {automated};
    \node [below=of auto label.west,anchor=west] (manual label) {manual};
    \draw [manual] (manual label -| hdl entry.west) -- (manual label);
    \draw [automatic] (auto label -| hdl entry.west) -- (auto label);
    \node [action, hw, below=14.5mm of hdl entry.west, anchor=west, draw=none, sharp corners] (hw label) {hardware-related};
    \node [action, sw, below=.5mm of hw label, draw=none, sharp corners] (sw label) {software-related};
  \end{tikzpicture}
  \caption{\workhandle{}'s automated \acrshort{hw}/\acrshort{sw} build and test flow.}
  \label{fig:build_and_test_flow}
\end{figure}
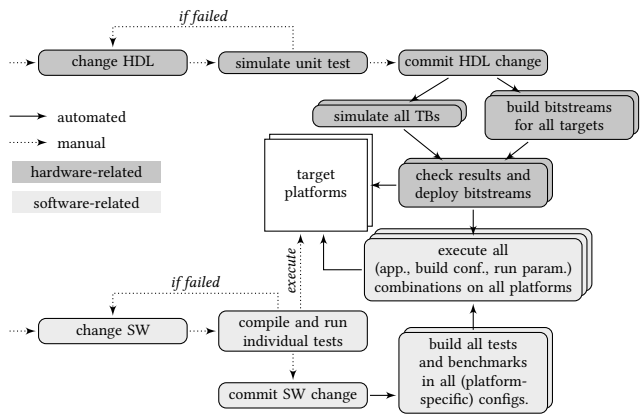

Automated full-sys\-tem builds and tests are a prerequisite for many effective development para\-digms. 
In our case, the system consists of an entire heterogeneous hardware and software stack.
As shown in Fig.~\ref{fig:build_and_test_flow}, different parts of the stack have to be built and tested depending on the change:
When modifying hardware, for example, we change code, evaluate the change with a unit test in the simulator, and commit the change once it passes that test.
Then, an integration server simulates all testbenches and builds bitstreams for all target platforms (in parallel) with the commit applied.
If, for a given target platform, all testbenches pass and the bitstream builds, the bitstream is deployed to the target platform.
Finally, the integration server starts all target platforms with updated bitstreams, runs all test and benchmark applications in all configurations (more on this below) on all platforms, collects the results, and reports them to the developer.


The automated hardware builds are relatively simple: the \gls{pmca} hardware description contains synthesis conditionals and parameters, whose values are defined in a platform-specific configuration file.
With the hardware description, this configuration file, and a uniform build script, \vivado{} generates platform-specific bitstreams.

The automated software builds and test runs are more complex:
%
To maximally optimize every \gls{pmca} kernel for streamlined execution, the \gls{pmca} runtime is co-compiled with each \gls{pmca} application and statically linked into a single binary with the same \emph{build parameters}.
There are platform-specific build parameters, which are defined in one configuration file per platform, and each application comes with its own set of build parameters and \emph{run arguments}, which are defined in an application-specific configuration file.
Each application must specify on which target platform which builds can be executed with which run arguments.
As the number of combinations grows exponentially with the number of parameters and arguments, listing all combinations manually would be redundant, error-prone work.
Instead, the combinations are specified in a compact, graph-based notation.
By flattening the graph, the integration server obtains the list of platform-application-parameter combinations, which are then built and executed automatically.
\section{Evaluation}
\label{sec:evaluation}

In this section, we describe the currently supported platforms and how their \gls{fpga} resources are used~(\secref{implemented_platforms}), demonstrate exploration of parallel execution and memory hierarchy usage~(\secref{parallelization_evaluation}), show the positive impact of \gls{svm} on the total \gls{pmca} run time~(\secref{SVM_evaluation}), and give examples how event tracing and analysis can be used to efficiently and accurately validate and characterize run-time behavior~(\secref{event_tracing_and_analysis_evaluation}).

\subsection{Supported Platforms}
\label{sec:implemented_platforms}

\workhandle{} is currently implemented on two different development platforms, and we are working on an implementation on the next-generation \xilinx{} \zynq{} \ultrascaleplus{} \glsknown{mpsoc}.
Porting \workhandle{} to a new \xilinx{} platform is an effort of approximately two man weeks.

\paragraph{\Glsunknown{juno}}
The \gls{juno} features an \arm{}v8-based, multi-cluster host \gls{cpu} (two A57 and four A53 cores), a Mali-T624 \gls{gpu}, and \SI{8}{\gibi\byte} of DDR3L \gls{dram}.
In addition, the \gls{soc} offers a low-latency \glsknown{axi} chip-to-chip interface (TLX-400) connecting to a \xilinx{} \virtex{} \gls{fpga}, through which \numrange{4}{8}~\gls{pmca} clusters on the \gls{fpga} can access the shared \gls{dram} coherently with the caches of the host.
The \arm{}v8 host \gls{cpu} runs 64b \linaro{} \linux{}~4.5 with a 64b root filesystem (both \texttt{aarch64-linux-gnu}) generated using the \openembedded{} build system.
We have configured the root filesystem to have multilib support, such that the host can also execute 32b binaries (\texttt{arm-linux-gnueabihf}) in \arm{}v7 mode, which guarantees compatibility of data and pointer types between the host and the 32b \gls{pmca} architecture in heterogeneous applications.\footnote{%
  This compatibility could also be achieved by running the application binary in ILP32 mode, which would allow the host to use \arm{}v8-specific \gls{cpu} features.
  However, the support for ILP32 is still experimental in \linaro{}.
}

\paragraph{\Glsunknown{zc706}}
The \xilinx{} \zynq{}-7045 \gls{soc} found on the \gls{zc706} combines an \arm{}v7, dual-core A9 host \gls{cpu} with a \kintex{} \gls{fpga} on a single chip.
The two subsystems are connected through a set of low-latency \glsknown{axi} interfaces and share \SI{1}{\gibi\byte} of DDR3 \gls{dram}.
Using the \gls{acp}, the single \gls{pmca} cluster instantiated in the \gls{fpga} can also coherently access data from the data caches of the host.
The main advantages of the \gls{zc706} is higher availability and better affordability compared to the \gls{juno}.
The 32b \arm{}v7 host \gls{cpu} runs \xilinx{} \linux{}~3.18 with a root filesystem generated using \buildroot{}.

\paragraph{\glsknown{fpga} resource utilization}
Tab.~\ref{tab:utilization} shows the \gls{fpga} resource utilization of the \gls{pmca} on the two development platforms in terms of \gls{lut} slices, \glspl{ff}, \gls{dsp} slices, and \gls{bram} cells.
The table lists both the absolute and the relative usage of the clusters and the top-level module containing also the host interfaces.
The configuration parameters selected for implementation are highlighted as bold and underlined text Tab.~\ref{tab:options} for the \gls{juno} and the \gls{zc706}, respectively.
The clusters dominate resource usage: \num{8} clusters on the \gls{juno} and \num{1} cluster on the \gls{zc706} account for more than 90\% and 80\% of the total resource usage, respectively.
While \gls{lut} and \gls{dsp} slices scale linearly from the single cluster on the \gls{zc706} to the \num{8} clusters on the \gls{juno}, \glspl{bram} and \glspl{ff} behave differently due to different instruction cache designs:
the larger, single-ported cache on the \gls{juno} uses more \glspl{ff} and less \glspl{bram} per cluster than the multi-ported cache on the \gls{zc706}.
Neither configuration includes \glspl{fpu}, and the integer data path alone uses relatively little \gls{dsp} slices, even though it supports multiplication and division.
The top-level configuration is identical for both platforms, with the exception of the different interfaces to the host and the number of clusters, which enlarges the registered \gls{soc} bus.
%
%
On both platforms, the available \glspl{lut} and \glspl{bram} are the limiting factors.
%
%
The \gls{pmca} can be clocked at \SI{31}{\mega\hertz} and \SI{57}{\mega\hertz} on the \gls{juno} and the \gls{zc706}, respectively.
The difference is due to the denser utilization of the \gls{juno} and the fact that the \virtex{} \gls{fpga} of the \gls{juno} consists of multiple dies connected through stacked silicon interconnects.

\begin{table}[h]
  \vspace*{\baselineskip}
  \centering
  \caption{\acrshort{pmca} \acrshort{fpga} resource utilization}
  \label{tab:utilization}
  \begin{threeparttable}
    \begingroup
    \setlength{\tabcolsep}{5pt} 
    \begin{tabular}{llrrrr}
    \toprule
      & & \multicolumn{2}{c}{\Glsknown{juno}} & \multicolumn{2}{c}{\Glsknown{zc706}} \\ \midrule
    \multirow{4}{*}{All Clusters}
      & LUT  & 936\,\si{\kilo} & 76\% & 128\,\si{\kilo} & 59\% \\
      & FF   & 450\,\si{\kilo} & 18\% &  43\,\si{\kilo} & 10\% \\
      & DSP  & 384             & 18\% &  48             &  5\% \\
      & BRAM & 1152            & 89\% & 384             & 70\% \\ \midrule
    \multirow{4}{*}{\shortstack{Top Level and\\Host Interface}}
      & LUT  &  70\,\si{\kilo} &  6\% &  24\,\si{\kilo} & 11\% \\
      & FF   &  61\,\si{\kilo} &  2\% &  26\,\si{\kilo} &  6\% \\
      & DSP  &   0             &  0\% &  0              &  0\% \\
      & BRAM &  75             &  6\% &  71             & 13\% \\
    \bottomrule
    \end{tabular}
  \endgroup
  \end{threeparttable}
  \vspace*{\baselineskip}
\end{table}

\subsection{Case Study: Parallel Speedup Analysis}
\label{sec:parallelization_evaluation}

\begin{figure}[ht]
  \centering
  \includegraphics[width=0.9\columnwidth]{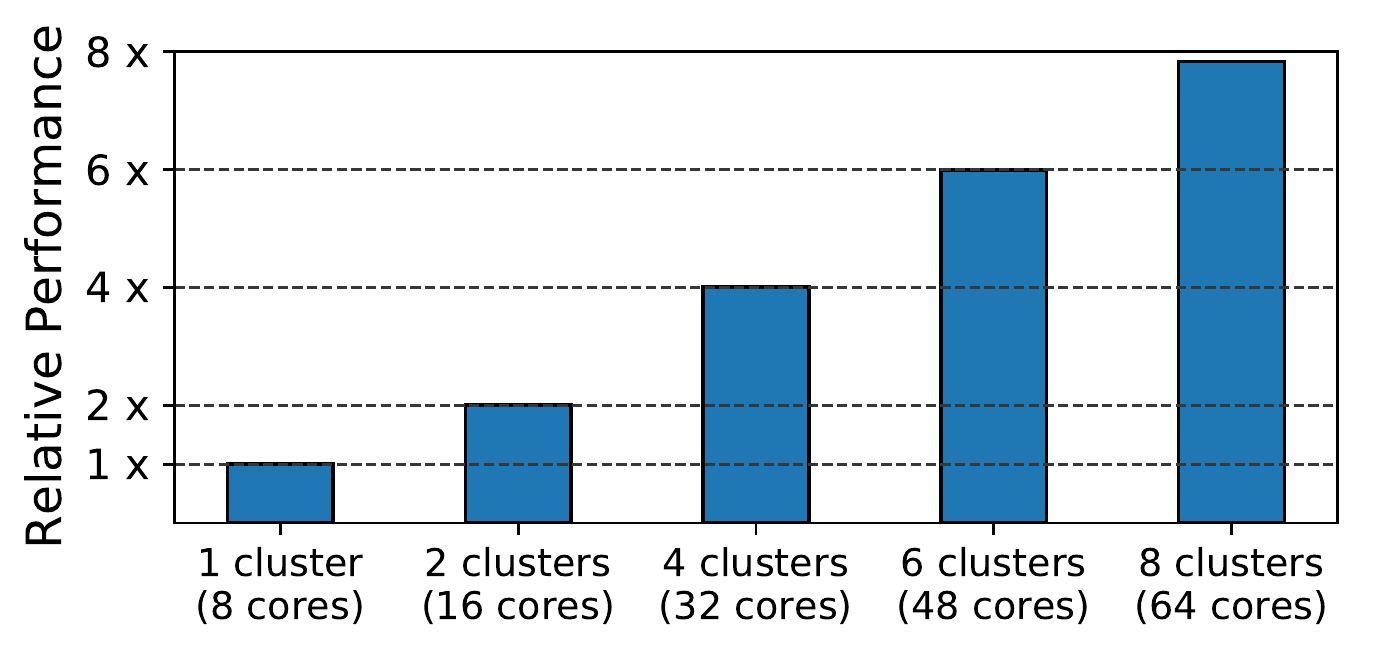}
  \caption{Overall execution speedup by parallelizing matrix-matrix multiplication.}
  \label{fig:MM_mul_exec_speedup}
\end{figure}

To demonstrate the parallel execution and data transfer capabilities of the \gls{pmca}, we use a matrix-matrix multiplication benchmark.
The computations for calculating the product of two matrices, \mbox{$ C = A B $}, are distributed over the clusters by tiling $ A $ and $ C $ row-wise.
Each cluster iterates over its rows and parallelizes each row block-wise over its cores: it transfers a row of $ A $ and a column of $ B $ from the \gls{dram} to its local L1 \gls{spm} banks, computes the resulting row of $ C $ into its L1 \gls{spm}, and transfers the resulting row to the \gls{dram}.

\Figref{MM_mul_exec_speedup} shows the speedup achieved when parallelizing the workload over multiple clusters.
In the baseline (leftmost bar), a single cluster performs the work.
The bars to the right of the baseline are for two, four, six, and eight clusters.
Parallelizing execution over two, four, and six clusters leads to ideal speedups compared to the baseline. 
For eight clusters, the interconnect between the clusters becomes the bottleneck to data transfers and limits the speedup to ca.\ 2\% below the ideal value.
In the evaluated implementation, the interconnect is a bus, which provides low latency but no scalable bandwidth.
A \acrlong{noc}, which is the other option for the interconnect between the clusters, scales in bandwidth and can thus, depending on the target workload, reduce the overall execution time by supporting parallel data transfers for even more \glspl{pe}.

\subsection{Case Study: Virtual Memory Performance Analysis}
\label{sec:SVM_evaluation}

\begin{figure}[ht]
  \centering
  \includegraphics[width=\columnwidth]{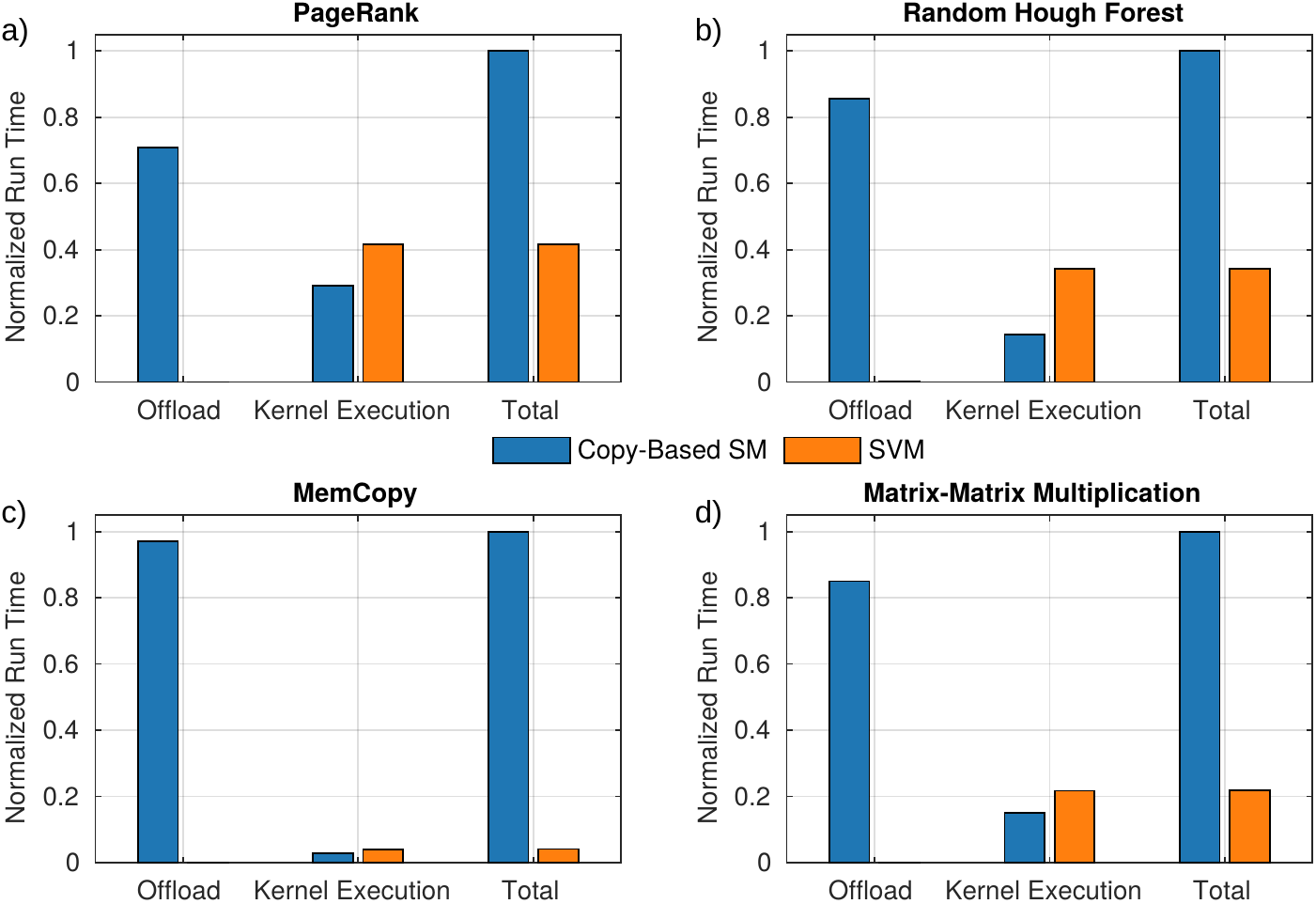}
  \caption{Offload and kernel execution time for different benchmarks with and without \gls{svm} support.}
  \label{fig:svm_run_time}
\end{figure}

\Gls{svm} support in the \gls{pmca} is essential for efficient data sharing between host and the \gls{pmca}:
Without \gls{svm}, data must be copied to and from a dedicated, physically-contiguous, uncached memory section before and after accelerator execution, respectively.
This copy operation depends on the data structure and may be very complex; e.g., the values of all pointers in a linked data structure must be changed.
With \gls{svm}, offloading simply means passing a pointer.

\Figref{svm_run_time} shows the run time of different benchmarks executed on the \gls{pmca} with (orange, right bar in a pair) and without \gls{svm} (blue, left bar in a pair).
The run time is broken down into offload time, i.e., the time it takes the host to offload the computation and prepare the data for the \gls{pmca}, and the actual kernel execution time on the \gls{pmca}.
All times are normalized to the total run time without \gls{svm}.
\textbf{PageRank~(a)} is a well-known algorithm for analyzing the connectivity of graphs and is used, e.g., for ranking web sites.
It is based on a \gls{lds}, which makes copy-based offloading expensive because the host must modify many pointers.
With \gls{svm}, virtual addresses must be translated at run time.
This causes a run time overhead if translations are not in the \gls{tlb} of the \gls{rab}.
Nonetheless, the offload time of copy-based SM dominates, and \gls{svm} reduces the run time by nearly 60\%.
\textbf{Random Hough Forests~(b)} consist of multiple binary decision trees and are used, e.g., for image classification.
The trees have a very large memory footprint, but only a part of them is accessed, depending on the input data.
With \gls{svm}, the \gls{pmca} can readily access the entire trees by performing the necessary address translations at run time.
With copy-based SM, the trees must be made available to the \gls{pmca} in their entirety before classification can start.
This leads to a lot of data being copied by the host that is never accessed by the \gls{pmca}.
\Gls{svm} reduces the run time by more than 60\%.
\textbf{MemCopy~(c)} simply copies a large array into the \gls{pmca} and back to memory.
This benchmark is representative of streaming applications that require the \gls{pmca} to perform only little work.
With copy-based SM, letting the host copy data to and from the physically contiguous, uncached memory to prepare the offload clearly dominates the run time.
In contrast, the \gls{pmca} benefits from high-bandwidth \gls{dma} transfers.
\Gls{svm} removes the need for data copying by the host, reducing the total run time by more than 95\%.
The \textbf{matrix-matrix multiplication benchmark~(d)} involves three matrices stored in arrays, thus shows the same basic behavior as MemCopy.
However, as the \gls{pmca} performs computations while traversing the data, the copy-based offload becomes a lesser part of the total run time.
In this case, \gls{svm} reduces the total run time by nearly 80\%.

\fontsize{8.8}{10.9}\selectfont
\subsection{Case Study: Advanced Event Tracing}
\label{sec:event_tracing_and_analysis_evaluation}

\begin{figure}
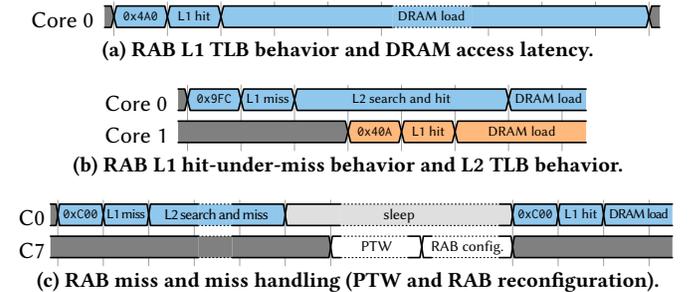

  \tikzstyle{timing style}=[
    timing/dslope=0.1,
    timing/.style={x=5ex,y=2ex},
    x=5ex,
    timing/rowdist=3ex,
    timing/coldist=1ex,
  ]
  \tikzstyle{ellipsis}=[thin, densely dotted]
  \tikzstyle{sig0}=[fill=sig0color]
  \tikzstyle{sig1}=[fill=sig1color]
  \tikzstyle{sleep}=[fill=lightgray!50!white]
  \tikzstyle{clockline}=[semitransparent, darkgray, very thin]
  \newcommand{\clockgrid}[1]{
    \begin{background}
      \vertlines[clockline]{#1}
    \end{background}
  }
  \begin{subfigure}{\columnwidth}
    \centering
    \begin{tikztimingtable}[timing style]
      Core 0  & .125U [sig0] D{\strut{}\texttt{0x4A0}} D{\strut{}L1 hit} 3.25D ;[ellipsis, sig0] 1.5D O{3.125D{\strut{}\acrshort{dram} load}} ;[sig0] 3.25D{} ; .25U \\
      \extracode\clockgrid{0.175,1.175,...,6,7.175,8.175,...,11}
    \end{tikztimingtable}
    \vspace*{-.5\baselineskip}
    \caption{\acrshort{rab} L1 \acrshort{tlb} behavior and \acrshort{dram} access latency.}
  \end{subfigure}
  \\[.25\baselineskip]
  \begin{subfigure}{\columnwidth}
    \centering
    \begin{tikztimingtable}[timing style]
      Core 0  & .125U [sig0] D{\strut{}\texttt{0x9FC}} D{\strut{}L1 miss} 4D{\strut{}L2 search and hit} 1.5D{\strut{}\acrshort{dram} load} \\
      Core 1  & 3.125U [sig1] D{\texttt{\strut{}0x40A}} D{\strut{}L1 hit} 2.5D{\strut{}\acrshort{dram} load} \\
      \extracode\clockgrid{0.175,1.175,...,8}
    \end{tikztimingtable}
    \vspace*{-.5\baselineskip}
    \caption{\acrshort{rab} L1 hit-under-miss behavior and L2 \acrshort{tlb} behavior.}
  \end{subfigure}
  \\[.25\baselineskip]
  \tikzstyle{overlabel}=[yshift=.9ex, font=\fontsize{5.7pt}{5.7pt}\selectfont]
  \begin{subfigure}{\columnwidth}
    \centering
    \begin{tikztimingtable}[%
      timing style,
      timing/dslope=.075,
      timing/.style={x=4.25ex,y=2ex},
      x=4.25ex,
      timing/coldist=.5ex,
    ]
      C0  & .125U [sig0] D{\strut{}\texttt{0xC00}} D{\strut{}\textls[-50]{L1 miss}} 1.125D ;[ellipsis,sig0] .75D N(p00) ;[sig0] 1.125D{} ; [sleep] 1.25D N(p01) ;[ellipsis,sleep] 3.5D ;[sleep] .25D{} ;[sig0] D{\strut{}\texttt{0xC00}} D{\strut{}L1 hit} 1.55D{\strut{}\textls[-75]{\acrshort{dram} load}}\\
      C7  & 3.25U [ellipsis] .75U; 2.125U .25D ;[ellipsis] 1.5D; N(p10) .25D{} .25D ;[ellipsis] 1.5D; N(p11) .25D 3.55U \\
      \extracode
      \clockgrid{0.1625,1.1625,...,7,10.1625,11.1625,...,14}
      \draw (p00) node[xshift=-1.5ex,overlabel] {\strut{}\textls[-50]{L2 search and miss}};
      \draw (p01) node[xshift=5.5ex,overlabel] {\strut{}sleep};
      \draw (p10) node[xshift=-3ex,overlabel] {\strut{}\acrshort{ptw}};
      \draw (p11) node[xshift=-3ex,overlabel] {\strut{}\acrshort{rab} config.};
    \end{tikztimingtable}
    \vspace*{-.5\baselineskip}
    \caption{\acrshort{rab} miss and miss handling (\acrshort{ptw} and \acrshort{rab} reconfiguration).}
  \end{subfigure}
  \caption{%
    Memory access events by different cores at the \acrshort{rab}.
    \normalfont\small
    For space efficiency, only the \acrshortpl{lsb} of each \acrshort{va} are shown and events that take many clock cycles to complete have been compressed, as indicated by dots.
  }
  \label{fig:memory_access_events}
\end{figure}

To evaluate different aspects of the interaction between \gls{pmca} and host, we inserted event tracers on the \glsknown{axi} read and write request and response channels of the \gls{rab} as well as on its configuration port.
We then executed different short programs on the \gls{pmca} to stimulate different behaviors of the memory subsystem.
The \gls{axi} tracers recorded raw memory access requests and responses and identified the related core through the \gls{axi} ID.
Finally, we used the event analysis tool to convert that data into series of events per core for each program, as shown in \figref{memory_access_events}.

In the first program~(a), a single core loads data from the \gls{dram} through a \gls{va} that is in the L1 \gls{tlb} of the \gls{rab}.
After a single-cycle address translation, the load is passed to the \gls{dram}.
With the \gls{pmca} running at much lower clock frequencies than it would in a silicon implementation, this load takes fewer cycles (an average of \num{7.8} at \SI{20}{\mega\hertz}) than it would in a real \gls{hesoc}, which would distort performance evaluations. 
By tracing all memory accesses in the execution of a benchmark, however, performance can be accurately determined by multiplying all access latencies with the implementation-to-emulation clock ratio.

In the second program~(b), one core accesses a \gls{va} that misses in the L1 \gls{tlb}, which triggers a multi-cycle search in the L2 \gls{tlb}.
While the L2 \gls{tlb} is being searched, a second core accesses a \gls{va} that is in the L1 \gls{tlb} and is indeed translated within a single clock cycle.
To verify that this hit-under-miss behavior is always maintained, the analyzer supports definable assertions.
Additionally, the number of clock cycles taken to find a \gls{va} in the L2 \gls{tlb} can be used to evaluate different placement strategies in the set-associative L2 \gls{tlb}.

In the third program~(c), a core accesses a \gls{va} that misses in both the L1 and the L2 \gls{tlb}, upon which it reports the miss to another core and goes to sleep.
The other core handles the miss through the \gls{vmm} library by walking the page table and configuring a L1 \gls{tlb} entry to translate that \gls{va}.
It then wakes the first core, which retries the memory access and proceeds with the load.
We used this to evaluate our \gls{vmm} implementation on the \gls{pmca} in~\cite{Vogel:2017:EfficientVirtualMemory}.
\section{Related Work}

\workhandle{} extends the principle of prototyping computer architectures on \glspl{fpga} to \glspl{hesoc}. 
In the \gls{fame} taxonomy~\cite{Tan:2010:FAME}, \workhandle{} is a Direct \gls{fame} system, meaning it implements the target architecture with a one-to-one correspondence in clock cycles on an \gls{fpga}.
More sophisticated \gls{fame} levels decouple timing and functionality, exchange structural equivalence for modeling abstractions, and share \gls{fpga} resources in time between components of the target architecture to increase model flexibility and emulation throughput.
An example of a sophisticated \gls{fame} system is \ramp{} Gold~\cite{Tan:2010:RAMPGold}, which is designed for the rapid early-design-space exploration of manycore systems.
It is cycle-accurate and comparable in throughput to \workhandle{}, but requires the development of a behavioral model that is not directly used in the silicon implementation.
In contrast to highly sophisticated \gls{fame} systems, \workhandle{} is not designed for early-stage design explorations but for the evaluation, advancement, and extension of a proven \gls{pmca} template and for studying the integration of \glspl{pmca} in a \gls{hesoc}.
By staying as close to the silicon implementation as possible, co-development and maintenance of separate models are avoided.
Commercial Direct \gls{fame} systems, such as \cadence{} \palladium{} and \mentorgraphics{} \veloce{}, are targeted at the verification of entire \glsknownpl{ic}.
To reach the required capacity, they employ custom logic simulation engines and highly intrusive tracing systems in addition to \glspl{fpga}.
They come with proprietary tools and protective licenses at very high costs, which bars the vast majority of the research community from using them.

The \gls{farm}~\cite{Oguntebi:2010:FARM} is a system for prototyping custom hardware implemented on an \gls{fpga} that is connected to an \amd{} multiprocessor.
Both \gls{farm} and \workhandle{} provide a cache-coherent link to the host processor and data transfer (or \gls{dma}) engines.
While \gls{farm} leaves the task of implementing an accelerator from scratch and integrating it with the system to the researcher, \workhandle{} comes with a \riscv{} manycore implementation, a heterogeneous toolchain, and tools to allow efficient hardware and software research using standard benchmarks and real-world applications.

\intel{} uses \glspl{fpga} to prototype heterogeneous---in their definition two sets of cores of the same \gls{isa} but different power-performance design points---architectures~\cite{Intel:2010:HybridProcessorEmulationOnFPGA,Intel:2012:QuickIA}.
They combine a \xeon{}~\cite{Intel:2010:HybridProcessorEmulationOnFPGA} and an \atom{}~\cite{Intel:2012:QuickIA} \gls{cpu} with an \gls{fpga} on which they implement up to four ``very old''~\cite{Intel:2012:QuickIA} \pentium{}~4 cores.
While an evaluation platform with a \xeon{} and an \atom{} \gls{cpu} (both hardwired) was shared with selected academic partners, the reconfigurable, \gls{fpga}-based prototypes remain restricted to \intel{}~\cite{Intel:2012:QuickIA}.
\workhandle{}, on the other hand, is openly available, implements a modern \riscv{} manycore on an \gls{fpga}, and uses the extended concept of heterogeneity with different \glspl{isa}.

\workhandle{} is more than a \gls{pmca} implemented on an \gls{fpga}, but its \gls{pmca} implementation is nonetheless related to the following recent works:
%
\openpiton{}~\cite{Balkind:2016:OpenPiton} is the first open-source, multithreaded manycore processor and is available in \gls{fpga} implementations for prototyping.
Our \gls{pmca} implementation on the \gls{fpga} differs from that of \openpiton{} in two ways:
First, our \gls{pmca} implements the \riscv{} \gls{isa}, which has recently gained a lot of momentum.
Second, it allows evaluation on the \gls{fpga} with more cores: we currently support up to 64 cores compared to \openpiton{}'s maximum of 4 cores (both on a \xilinx{} \virtex{}, albeit of different size).
%
GRVI Phalanx~\cite{Gray:2016:GRVIPhalanx} is an array of clusters of \riscv{} cores interconnected by a \gls{noc}.
Cores, clusters, and the \gls{noc} are optimized for \glspl{fpga} and utilize \gls{fpga} blocks very efficiently, allowing to implement hundreds of RV32I cores on a mid-range \gls{fpga}.
While \glspl{fpga} are \emph{the} design target of GRVI Phalanx, \workhandle{} uses \glspl{fpga} as a prototyping target to support a wide range of implementation targets and architectural exploration.
Moreover, GRVI Phalanx is programmed bare-metal, whereas the \gls{pmca} on \workhandle{} comes with a runtime that supports well-established programming paradigms such as \openmp{} including seamless accelerator integration.
%
\lowrisc{}~\cite{Bradbury:2014:lowRISC} is a work-in-progress open-source \glsknown{soc} implementing the \riscv{} \gls{isa}.
Its goal is to lower the barrier of entry to producing custom silicon by establishing an ecosystem of \glsknown{ip} blocks around \riscv{} cores.
In contrast, \workhandle{} aims to facilitate exploration on all layers of software and hardware in \glspl{hesoc} by implementing a modifiable, working full-stack prototype accompanied by tools for validation and evaluation of novel concepts.

\section{Conclusion}

We presented \workhandle{}, the first heterogeneous manycore research platform, which unites an \arm{} \cortexa{} host processor with a fully modifiable \riscv{} manycore implemented on an \gls{fpga}.
Our heterogeneous software stack, which supports \gls{svm} and \openmp{}~4.5, tremendously simplifies porting of standard benchmarks and real-world applications, thereby enabling system-level research.
Our profiling and automated verification solutions enable efficient hardware and software research on all layers.
We have been successfully using \workhandle{} in our research over the last years and will continue its development.
To further advance the research community, we are currently working towards releasing \workhandle{} under an open-source license on \href{http://pulp-platform.org/hero}{\texttt{pulp-platform.org/hero}}.

\renewcommand{\bibliofont}{\fontsize{9.2}{10}\selectfont}
\setlength{\bibsep}{.3ex plus .5ex}
\bibliographystyle{ACM-Reference-Format}
\bibliography{paper}


\begin{thebibliography}{00}


\ifx \showCODEN    \undefined \def \showCODEN     #1{\unskip}     \fi
\ifx \showDOI      \undefined \def \showDOI       #1{#1}\fi
\ifx \showISBNx    \undefined \def \showISBNx     #1{\unskip}     \fi
\ifx \showISBNxiii \undefined \def \showISBNxiii  #1{\unskip}     \fi
\ifx \showISSN     \undefined \def \showISSN      #1{\unskip}     \fi
\ifx \showLCCN     \undefined \def \showLCCN      #1{\unskip}     \fi
\ifx \shownote     \undefined \def \shownote      #1{#1}          \fi
\ifx \showarticletitle \undefined \def \showarticletitle #1{#1}   \fi
\ifx \showURL      \undefined \def \showURL       {\relax}        \fi
\providecommand\bibfield[2]{#2}
\providecommand\bibinfo[2]{#2}
\providecommand\natexlab[1]{#1}
\providecommand\showeprint[2][]{arXiv:#2}

\bibitem[\protect\citeauthoryear{Balkind et~al\mbox{.}}{Balkind
  et~al\mbox{.}}{2016}]%
        {Balkind:2016:OpenPiton}
\bibfield{author}{\bibinfo{person}{J. Balkind} \textit{et~al\mbox{.}}}
  \bibinfo{year}{2016}\natexlab{}.
\newblock \showarticletitle{{OpenPiton}: An Open Source Manycore Research
  Framework}. In \bibinfo{booktitle}{{\em Proceedings of the Twenty-First
  International Conference on Architectural Support for Programming Languages
  and Operating Systems}} {\em (\bibinfo{series}{ASPLOS '16})}.
\newblock


\bibitem[\protect\citeauthoryear{Binkert et~al\mbox{.}}{Binkert
  et~al\mbox{.}}{2011}]%
        {Binkert:2011:gem5}
\bibfield{author}{\bibinfo{person}{N. Binkert} \textit{et~al\mbox{.}}}
  \bibinfo{year}{2011}\natexlab{}.
\newblock \showarticletitle{The {gem5} Simulator}.
\newblock \bibinfo{journal}{{\em SIGARCH Comput. Archit. News\/}}
  \bibinfo{volume}{39}, \bibinfo{number}{2}, \bibinfo{pages}{1--7}.
\newblock
\showISSN{0163-5964}
\showDOI{%
\url{https://doi.org/10.1145/2024716.2024718}}


\bibitem[\protect\citeauthoryear{Bohnenstiehl et~al\mbox{.}}{Bohnenstiehl
  et~al\mbox{.}}{2017}]%
        {Bohnenstiehl:2017:KiloCore}
\bibfield{author}{\bibinfo{person}{B. Bohnenstiehl} \textit{et~al\mbox{.}}}
  \bibinfo{year}{2017}\natexlab{}.
\newblock \showarticletitle{{KiloCore}: A 32-nm 1000-Processor Computational
  Array}.
\newblock \bibinfo{journal}{{\em {IEEE} Journal of Solid-State Circuits
  ({JSSC})\/}} \bibinfo{volume}{52}, \bibinfo{number}{4},
  \bibinfo{pages}{891--902}.
\newblock


\bibitem[\protect\citeauthoryear{Bortolotti et~al\mbox{.}}{Bortolotti
  et~al\mbox{.}}{2016}]%
        {Bortolotti:2016:VirtualSoC}
\bibfield{author}{\bibinfo{person}{D. Bortolotti} \textit{et~al\mbox{.}}}
  \bibinfo{year}{2016}\natexlab{}.
\newblock \showarticletitle{VirtualSoC: A Research Tool for Modern MPSoCs}.
\newblock \bibinfo{journal}{{\em ACM Trans. Embed. Comput. Syst.\/}}
  \bibinfo{volume}{16}, \bibinfo{number}{1}, \bibinfo{numpages}{27}~pages.
\newblock
\showISSN{1539-9087}
\showDOI{%
\url{https://doi.org/10.1145/2930665}}


\bibitem[\protect\citeauthoryear{Bradbury et~al\mbox{.}}{Bradbury
  et~al\mbox{.}}{2014}]%
        {Bradbury:2014:lowRISC}
\bibfield{author}{\bibinfo{person}{A. Bradbury} \textit{et~al\mbox{.}}}
  \bibinfo{year}{2014}\natexlab{}.
\newblock \bibinfo{title}{Tagged Memory and Minion Cores in the {lowRISC}
  {SoC}}.
\newblock
\newblock
\showURL{%
\url{http://www.lowrisc.org/downloads/lowRISC-memo-2014-001.pdf}}


\bibitem[\protect\citeauthoryear{Butko et~al\mbox{.}}{Butko
  et~al\mbox{.}}{2016}]%
        {Butko:2016:bigLITTLESimulation}
\bibfield{author}{\bibinfo{person}{A. Butko} \textit{et~al\mbox{.}}}
  \bibinfo{year}{2016}\natexlab{}.
\newblock \showarticletitle{Full-System Simulation of big.LITTLE Multicore
  Architecture for Performance and Energy Exploration}. In
  \bibinfo{booktitle}{{\em IEEE MCSOC}}. \bibinfo{pages}{201--208}.
\newblock
\showDOI{%
\url{https://doi.org/10.1109/MCSoC.2016.20}}


\bibitem[\protect\citeauthoryear{Capotondi and Marongiu}{Capotondi and
  Marongiu}{2017}]%
        {Capotondi:2017:OpenMPOffloading}
\bibfield{author}{\bibinfo{person}{A. Capotondi} {and} \bibinfo{person}{A.
  Marongiu}.} \bibinfo{year}{2017}\natexlab{}.
\newblock \showarticletitle{Enabling Zero-copy OpenMP Offloading on the PULP
  Many-core Accelerator}. In \bibinfo{booktitle}{{\em Proceedings of the 20th
  International Workshop on Software and Compilers for Embedded Systems}} {\em
  (\bibinfo{series}{SCOPES '17})}. \bibinfo{publisher}{ACM},
  \bibinfo{address}{New York, NY, USA}, \bibinfo{pages}{68--71}.
\newblock
\showISBNx{978-1-4503-5039-6}
\showDOI{%
\url{https://doi.org/10.1145/3078659.3079071}}


\bibitem[\protect\citeauthoryear{Celio et~al\mbox{.}}{Celio
  et~al\mbox{.}}{2015}]%
        {Celio:2015:Boom}
\bibfield{author}{\bibinfo{person}{C. Celio} \textit{et~al\mbox{.}}}
  \bibinfo{year}{2015}\natexlab{}.
\newblock \bibinfo{title}{The Berkeley Out-of-Order Machine (BOOM): An
  Industry-Competitive, Synthesizable, Parameterized RISC-V Processor}.
\newblock
\newblock
\showURL{%
\url{http://www2.eecs.berkeley.edu/Pubs/TechRpts/2015/EECS-2015-167.html}}


\bibitem[\protect\citeauthoryear{Chen et~al\mbox{.}}{Chen
  et~al\mbox{.}}{2015}]%
        {Chen:2015:NnAccelerator}
\bibfield{author}{\bibinfo{person}{T. Chen} \textit{et~al\mbox{.}}}
  \bibinfo{year}{2015}\natexlab{}.
\newblock \showarticletitle{A High-Throughput Neural Network Accelerator}.
\newblock \bibinfo{journal}{{\em IEEE Micro\/}} \bibinfo{volume}{35},
  \bibinfo{number}{3}, \bibinfo{pages}{24--32}.
\newblock
\showISSN{0272-1732}
\showDOI{%
\url{https://doi.org/10.1109/MM.2015.41}}


\bibitem[\protect\citeauthoryear{Chitlur et~al\mbox{.}}{Chitlur
  et~al\mbox{.}}{2012}]%
        {Intel:2012:QuickIA}
\bibfield{author}{\bibinfo{person}{N. Chitlur} \textit{et~al\mbox{.}}}
  \bibinfo{year}{2012}\natexlab{}.
\newblock \showarticletitle{{QuickIA}: Exploring Heterogeneous Architectures on
  Real Prototypes}. In \bibinfo{booktitle}{{\em IEEE HPCA}}.
  \bibinfo{pages}{1--8}.
\newblock
\showISSN{1530-0897}
\showDOI{%
\url{https://doi.org/10.1109/HPCA.2012.6169046}}


\bibitem[\protect\citeauthoryear{de~Dinechin et~al\mbox{.}}{de~Dinechin
  et~al\mbox{.}}{2013}]%
        {DeDinechin:2013:KalrayMPPA}
\bibfield{author}{\bibinfo{person}{B.~D. de Dinechin} \textit{et~al\mbox{.}}}
  \bibinfo{year}{2013}\natexlab{}.
\newblock \showarticletitle{A Clustered Manycore Processor Architecture for
  Embedded and Accelerated Applications}. In \bibinfo{booktitle}{{\em IEEE
  HPEC}}. \bibinfo{pages}{1--6}.
\newblock
\showDOI{%
\url{https://doi.org/10.1109/HPEC.2013.6670342}}


\bibitem[\protect\citeauthoryear{Gautschi et~al\mbox{.}}{Gautschi
  et~al\mbox{.}}{2017}]%
        {Gautschi:2017:NtvRiscvDspExtensions}
\bibfield{author}{\bibinfo{person}{M. Gautschi} \textit{et~al\mbox{.}}}
  \bibinfo{year}{2017}\natexlab{}.
\newblock \showarticletitle{Near-Threshold RISC-V Core With DSP Extensions for
  Scalable IoT Endpoint Devices}.
\newblock \bibinfo{journal}{{\em IEEE TVLSI\/}} \bibinfo{volume}{PP},
  \bibinfo{number}{99}, \bibinfo{pages}{1--14}.
\newblock
\showISSN{1063-8210}
\showDOI{%
\url{https://doi.org/10.1109/TVLSI.2017.2654506}}


\bibitem[\protect\citeauthoryear{Gray}{Gray}{2016}]%
        {Gray:2016:GRVIPhalanx}
\bibfield{author}{\bibinfo{person}{J. Gray}.} \bibinfo{year}{2016}\natexlab{}.
\newblock \showarticletitle{{GRVI Phalanx}: A Massively Parallel {RISC-V}
  {FPGA} Accelerator Accelerator}. In \bibinfo{booktitle}{{\em 2016 IEEE 24th
  Annual International Symposium on Field-Programmable Custom Computing
  Machines (FCCM)}}. \bibinfo{pages}{17--20}.
\newblock
\showDOI{%
\url{https://doi.org/10.1109/FCCM.2016.12}}


\bibitem[\protect\citeauthoryear{Ho}{Ho}{2016}]%
        {AnandTech:2017:NvidiaTegraParker}
\bibfield{author}{\bibinfo{person}{J. Ho}.} \bibinfo{year}{2016}\natexlab{}.
\newblock \showarticletitle{NVIDIA's {Tegra Parker SoC}}.
\newblock \bibinfo{journal}{{\em AnandTech\/}}.
\newblock
\showURL{%
\url{http://www.anandtech.com/show/10596}}


\bibitem[\protect\citeauthoryear{Johnson et~al\mbox{.}}{Johnson
  et~al\mbox{.}}{2011}]%
        {Johnson:2011:Rigel}
\bibfield{author}{\bibinfo{person}{D. Johnson} \textit{et~al\mbox{.}}}
  \bibinfo{year}{2011}\natexlab{}.
\newblock \showarticletitle{Rigel: A 1,024-Core Single-Chip Accelerator
  Architecture}.
\newblock \bibinfo{journal}{{\em IEEE Micro\/}} \bibinfo{volume}{31},
  \bibinfo{number}{4}, \bibinfo{pages}{30--41}.
\newblock
\showISSN{0272-1732}
\showDOI{%
\url{https://doi.org/10.1109/MM.2011.40}}


\bibitem[\protect\citeauthoryear{Kanter}{Kanter}{2016}]%
        {MicroprocessorReport:2016:Riscv}
\bibfield{author}{\bibinfo{person}{D. Kanter}.}
  \bibinfo{year}{2016}\natexlab{}.
\newblock \showarticletitle{{RISC-V} Offers Simple, Modular {ISA}: New {CPU}
  Instruction Set Is Open and Extensible}.
\newblock
\showURL{%
\url{https://riscv.org/2016/04/risc-v-offers-simple-modular-isa}}


\bibitem[\protect\citeauthoryear{Lee et~al\mbox{.}}{Lee et~al\mbox{.}}{2016}]%
        {Lee:2016:AgileHardware}
\bibfield{author}{\bibinfo{person}{Y. Lee} \textit{et~al\mbox{.}}}
  \bibinfo{year}{2016}\natexlab{}.
\newblock \showarticletitle{An Agile Approach to Building RISC-V
  Microprocessors}.
\newblock \bibinfo{journal}{{\em IEEE Micro\/}} \bibinfo{volume}{36},
  \bibinfo{number}{2}, \bibinfo{pages}{8--20}.
\newblock
\showISSN{0272-1732}
\showDOI{%
\url{https://doi.org/10.1109/MM.2016.11}}


\bibitem[\protect\citeauthoryear{Li et~al\mbox{.}}{Li et~al\mbox{.}}{2009}]%
        {Li:2009:McPAT}
\bibfield{author}{\bibinfo{person}{S. Li} \textit{et~al\mbox{.}}}
  \bibinfo{year}{2009}\natexlab{}.
\newblock \showarticletitle{{McPAT}: An integrated power, area, and timing
  modeling framework for multicore and manycore architectures}. In
  \bibinfo{booktitle}{{\em IEEE/ACM MICRO}}. \bibinfo{pages}{469--480}.
\newblock
\showISSN{1072-4451}


\bibitem[\protect\citeauthoryear{Melpignano et~al\mbox{.}}{Melpignano
  et~al\mbox{.}}{2012}]%
        {Melpignano:2012:P2012}
\bibfield{author}{\bibinfo{person}{D. Melpignano} \textit{et~al\mbox{.}}}
  \bibinfo{year}{2012}\natexlab{}.
\newblock \showarticletitle{{Platform 2012}, a Many-core Computing Accelerator
  for Embedded SoCs: Performance Evaluation of Visual Analytics Applications}.
  In \bibinfo{booktitle}{{\em Proceedings of the 49th Annual Design Automation
  Conference}} {\em (\bibinfo{series}{DAC '12})}. \bibinfo{publisher}{ACM},
  \bibinfo{address}{New York, NY, USA}, \bibinfo{pages}{1137--1142}.
\newblock
\showISBNx{978-1-4503-1199-1}
\showDOI{%
\url{https://doi.org/10.1145/2228360.2228568}}


\bibitem[\protect\citeauthoryear{Oguntebi et~al\mbox{.}}{Oguntebi
  et~al\mbox{.}}{2010}]%
        {Oguntebi:2010:FARM}
\bibfield{author}{\bibinfo{person}{T. Oguntebi} \textit{et~al\mbox{.}}}
  \bibinfo{year}{2010}\natexlab{}.
\newblock \showarticletitle{{FARM}: A Prototyping Environment for
  Tightly-Coupled, Heterogeneous Architectures}. In \bibinfo{booktitle}{{\em
  IEEE FCCM}}. \bibinfo{pages}{221--228}.
\newblock
\showDOI{%
\url{https://doi.org/10.1109/FCCM.2010.41}}


\bibitem[\protect\citeauthoryear{Olofsson}{Olofsson}{2016}]%
        {Olofsson:2016:EpiphanyV}
\bibfield{author}{\bibinfo{person}{A. Olofsson}.}
  \bibinfo{year}{2016}\natexlab{}.
\newblock \showarticletitle{{Epiphany-V}: A 1024 processor 64-bit {RISC}
  System-On-Chip}.
\newblock \bibinfo{journal}{{\em CoRR\/}}  \bibinfo{volume}{abs/1610.01832}.
\newblock
\showURL{%
\url{http://arxiv.org/abs/1610.01832}}


\bibitem[\protect\citeauthoryear{Rossi et~al\mbox{.}}{Rossi
  et~al\mbox{.}}{2014}]%
        {Rossi:2014:PULP}
\bibfield{author}{\bibinfo{person}{D. Rossi} \textit{et~al\mbox{.}}}
  \bibinfo{year}{2014}\natexlab{}.
\newblock \showarticletitle{Energy efficient parallel computing on the {PULP}
  platform with support for {OpenMP}}. In \bibinfo{booktitle}{{\em IEEEI}}.
  \bibinfo{pages}{1--5}.
\newblock
\showDOI{%
\url{https://doi.org/10.1109/EEEI.2014.7005803}}


\bibitem[\protect\citeauthoryear{Smith}{Smith}{2017a}]%
        {Anandtech:2017:AppleA10X}
\bibfield{author}{\bibinfo{person}{R. Smith}.}
  \bibinfo{year}{2017}\natexlab{a}.
\newblock \showarticletitle{Apple's {A10X SoC}}.
\newblock \bibinfo{journal}{{\em AnandTech\/}}.
\newblock
\showURL{%
\url{http://www.anandtech.com/show/11596}}


\bibitem[\protect\citeauthoryear{Smith}{Smith}{2017b}]%
        {AnandTech:2017:SamsungExynos8895}
\bibfield{author}{\bibinfo{person}{R. Smith}.}
  \bibinfo{year}{2017}\natexlab{b}.
\newblock \showarticletitle{Samsung's {Exynos 8895 SoC}}.
\newblock \bibinfo{journal}{{\em AnandTech\/}}.
\newblock
\showURL{%
\url{http://www.anandtech.com/show/11149}}


\bibitem[\protect\citeauthoryear{Tan et~al\mbox{.}}{Tan et~al\mbox{.}}{2010a}]%
        {Tan:2010:FAME}
\bibfield{author}{\bibinfo{person}{Z. Tan} \textit{et~al\mbox{.}}}
  \bibinfo{year}{2010}\natexlab{a}.
\newblock \showarticletitle{A Case for {FAME}: {FPGA} Architecture Model
  Execution}. In \bibinfo{booktitle}{{\em Proceedings of the 37th Annual
  International Symposium on Computer Architecture}} {\em
  (\bibinfo{series}{ISCA '10})}. \bibinfo{publisher}{ACM},
  \bibinfo{address}{New York, NY, USA}, \bibinfo{pages}{290--301}.
\newblock
\showISBNx{978-1-4503-0053-7}
\showDOI{%
\url{https://doi.org/10.1145/1815961.1815999}}


\bibitem[\protect\citeauthoryear{Tan et~al\mbox{.}}{Tan et~al\mbox{.}}{2010b}]%
        {Tan:2010:RAMPGold}
\bibfield{author}{\bibinfo{person}{Z. Tan} \textit{et~al\mbox{.}}}
  \bibinfo{year}{2010}\natexlab{b}.
\newblock \showarticletitle{{RAMP Gold}: An {FPGA}-based Architecture Simulator
  for Multiprocessors}. In \bibinfo{booktitle}{{\em Proceedings of the 47th
  Design Automation Conference}} {\em (\bibinfo{series}{DAC '10})}.
  \bibinfo{publisher}{ACM}, \bibinfo{address}{New York, NY, USA},
  \bibinfo{pages}{463--468}.
\newblock
\showISBNx{978-1-4503-0002-5}
\showDOI{%
\url{https://doi.org/10.1145/1837274.1837390}}


\bibitem[\protect\citeauthoryear{Traber et~al\mbox{.}}{Traber
  et~al\mbox{.}}{2016}]%
        {Traber:2016:Pulpino}
\bibfield{author}{\bibinfo{person}{A. Traber} \textit{et~al\mbox{.}}}
  \bibinfo{year}{2016}\natexlab{}.
\newblock \bibinfo{title}{{PULPino}: A small single-core {RISC-V SoC}}.
\newblock
\newblock
\showURL{%
\url{https://riscv.org/wp-content/uploads/2016/01/Wed1315-PULP-riscv3_noanim.pdf}}


\bibitem[\protect\citeauthoryear{Vogel et~al\mbox{.}}{Vogel
  et~al\mbox{.}}{2015}]%
        {Vogel:2015:RAB}
\bibfield{author}{\bibinfo{person}{P. Vogel} \textit{et~al\mbox{.}}}
  \bibinfo{year}{2015}\natexlab{}.
\newblock \showarticletitle{Lightweight Virtual Memory Support for Many-core
  Accelerators in Heterogeneous Embedded {SoCs}}.
\newblock \bibinfo{pages}{45--54}.
\newblock
\showISBNx{978-1-4673-8321-9}
\showURL{%
\url{http://dl.acm.org/citation.cfm?id=2830840.2830846}}


\bibitem[\protect\citeauthoryear{Vogel et~al\mbox{.}}{Vogel
  et~al\mbox{.}}{2016}]%
        {Vogel:2016:PointerRich}
\bibfield{author}{\bibinfo{person}{P. Vogel} \textit{et~al\mbox{.}}}
  \bibinfo{year}{2016}\natexlab{}.
\newblock \showarticletitle{Lightwight Virtual Memory Support for Zero-Copy
  Sharing of Pointer-Rich Data Structures in Heterogeneous Embedded SoCs}.
\newblock \bibinfo{journal}{{\em IEEE TPDS\/}} \bibinfo{volume}{28},
  \bibinfo{number}{7}, \bibinfo{pages}{1947--1959}.
\newblock


\bibitem[\protect\citeauthoryear{Vogel et~al\mbox{.}}{Vogel
  et~al\mbox{.}}{2017}]%
        {Vogel:2017:EfficientVirtualMemory}
\bibfield{author}{\bibinfo{person}{P. Vogel} \textit{et~al\mbox{.}}}
  \bibinfo{year}{2017}\natexlab{}.
\newblock \showarticletitle{Efficient Virtual Memory Sharing via On-Accelerator
  Page Table Walking in Heterogeneous Embedded SoCs}.
\newblock \bibinfo{journal}{{\em ACM TECS\/}} \bibinfo{volume}{PP},
  \bibinfo{number}{99}, \bibinfo{pages}{1--19}.
\newblock


\bibitem[\protect\citeauthoryear{Wang et~al\mbox{.}}{Wang
  et~al\mbox{.}}{2010}]%
        {Intel:2010:HybridProcessorEmulationOnFPGA}
\bibfield{author}{\bibinfo{person}{Q. Wang} \textit{et~al\mbox{.}}}
  \bibinfo{year}{2010}\natexlab{}.
\newblock \showarticletitle{An {FPGA} Based Hybrid Processor Emulation
  Platform}. In \bibinfo{booktitle}{{\em FPL}}. \bibinfo{pages}{25--30}.
\newblock
\showISSN{1946-147X}
\showDOI{%
\url{https://doi.org/10.1109/FPL.2010.16}}


\bibitem[\protect\citeauthoryear{Waterman}{Waterman}{2016}]%
        {Waterman:2016:RiscvIsa}
\bibfield{author}{\bibinfo{person}{A. Waterman}.}
  \bibinfo{year}{2016}\natexlab{}.
\newblock \bibinfo{title}{Design of the RISC-V Instruction Set Architecture}.
\newblock
\newblock
\showURL{%
\url{http://www2.eecs.berkeley.edu/Pubs/TechRpts/2016/EECS-2016-1.html}}


\bibitem[\protect\citeauthoryear{Zimmer et~al\mbox{.}}{Zimmer
  et~al\mbox{.}}{2016}]%
        {Zimmer:2016:RiscvVectorProcessor}
\bibfield{author}{\bibinfo{person}{B. Zimmer} \textit{et~al\mbox{.}}}
  \bibinfo{year}{2016}\natexlab{}.
\newblock \showarticletitle{A {RISC-V} Vector Processor With
  Simultaneous-Switching Switched-Capacitor {DC-DC} Converters in 28 nm
  {FDSOI}}.
\newblock \bibinfo{journal}{{\em IEEE JSSC\/}} \bibinfo{volume}{51},
  \bibinfo{number}{4}, \bibinfo{pages}{930--942}.
\newblock
\showISSN{0018-9200}
\showDOI{%
\url{https://doi.org/10.1109/JSSC.2016.2519386}}


\end{thebibliography}

\end{document}